\documentclass[english,journal=jctcce,manuscript=article]{achemso}
\usepackage[T1]{fontenc}
\usepackage[latin9]{inputenc}
\usepackage{amsmath}
\usepackage{graphicx}

\makeatletter

\title{Stochastic self-consistent second-order Green's function method for
correlation energies of large electronic systems}
\author{Daniel Neuhauser}
\email{dxn@ucla.edu}
\affiliation{Department of Chemistry and Biochemistry, University of California
at Los Angeles, California 90095, USA}
\author{Roi Baer}
\email{roi.baer@huji.ac.il}
\affiliation{Fritz Haber Center for Molecular Dynamics, Institute of Chemistry,
The Hebrew University of Jerusalem, Jerusalem 91904, Israel}
\author{Dominika Zgid}
\email{dominika.zgid@gmail.com}
\affiliation{Department of Chemistry, University of Michigan, Ann Arbor, Michigan
48109, USA}
\providecommand{\tabularnewline}{\\}

\makeatother

\usepackage{babel}
\begin{document}
\begin{abstract}
The second-order Matsubara Green's function method (GF2) is a robust
temperature dependent quantum chemistry approach, extending beyond
the random-phase approximation. However, till now the scope of GF2
applications was quite limited as they require computer resources
which rise steeply with system size. In each step of the self-consistent
GF2 calculation there are two parts: the estimation of the self-energy
from the previous step's Green's function, and updating the Green's
function from the self-energy. The first part formally scales as the
fifth power of the system size while the second has a much gentler
cubic scaling. Here, we develop a stochastic approach to GF2 (sGF2)
which reduces the fifth power scaling of the first step to merely
quadratic, leaving the overall sGF2 scaling as cubic. We apply the
method to linear hydrogen chains containing up to 1000 electrons,
showing that the approach is numerically stable, efficient and accurate.
The stochastic errors are very small, of the order of 0.1\% or less
of the correlation energy for large systems, with only a moderate
computational effort. The first iteration of GF2 is an MP2 calculation
that is done in \emph{linear scaling}, hence we obtain an extremely
fast stochastic MP2 (sMP2) method as a by-product. While here we consider
finite systems with large band gaps where at low temperatures effects
are negligible, the sGF2 formalism is temperature dependent and general
and can be applied to finite or periodic systems with small gaps at
finite temperatures.
\end{abstract}

\section{Introduction}

Second-order Green's function (GF2) is a temperature-dependent self-consistent
perturbation approach where the Green's function is iteratively renormalized.
At self-consistency the self-energy which accounts for the many-body
correlation effects is a functional of the Green's function, $\Sigma(G)$.
The GF2 approximation as implemented here is described by the diagrams
in Fig.~\ref{Fig:SE_GF2} and employs Matsubara Green\textquoteright s
functions that are temperature dependent and expressed on the imaginary
axis.\cite{baym1962self,Dahlen05,Phillips2014,Phillips2015} The implementation
we discuss, for total energies, relies on thermal Matsubara Green's
functions instead of real time Green's functions.\cite{Hedin1965,Fetter1971,Onida2002}
This offers advantages in terms of stability and smoothness of the
self-energy. 

Upon convergence the GF2 method includes all second order skeleton
diagrams dressed with the renormalized second order Green's function
propagators, as illustrated in Fig.~\ref{Fig:SE_GF2}. Specifically,
as shown in Ref.~\citenum{Phillips2014}, GF2, which at convergence
is reference independent, preserves the desirable features of M�ller-Plesset
perturbation theory (MP2) while avoiding the divergences that appear
when static correlation is important. Additionally, GF2 possesses
only a very small fractional charge and spin error,\cite{Phillips2015}
less than either typical hybrid density functionals or RPA with exchange,
therefore having a minimal many-body self-interaction error. In solids
GF2 describes the insulating and Mott regimes and recovers the internal
and free energy for multiple solid phases.\cite{Rusakov2016,Welden16}
Moreover, GF2 is useful for efficient Green's function embedding techniques
such as in the self-energy embedding method (SEET).\cite{Kananenka2015,Tran15b,Tran16,Tran17,zgid_njp17}

The formal advantages of GF2 come, however, with a price tag. The
calculation of the self-energy matrix scales as ${\rm O}(n_{\tau}N^{5})$,
where $n_{\tau}$ is the size of the imaginary time grid and $N$
the number of atomic orbitals (AOs). This leads to steep numerical
costs which prevent application of GF2 to systems larger than a few
dozen electrons. The application to larger systems requires therefore
a different paradigm and here we therefore develop a statistical formulation
of GF2 that calculates the self-energy matrix in linear-scaling.

The key to the present development, distinguishing it from previous
work \cite{Thom2007a,kozik2010diagrammatic,willow2013stochastic},
is the conversion of nested summations into stochastic averages. Our
method draws from previous work on stochastic electronic structure
methods, including stochastic- density functional theory (sDFT),\cite{Baer2013,Neuhauser2014a,arnon2017equilibrium},
sDFT with long-range exact exchange,\cite{Baer2012} multi-exciton
generation,\cite{Baer2012a} Moller-Plesset perturbation theory (sMP2),\cite{Neuhauser2013,Ge2013,takeshita2017stochastic}
random-phase approximation (sRPA),\cite{Neuhauser2013a} GW approximation
(sGW),\cite{Neuhauser2014,Vlcek2017,vlcek2017self} time-dependent
DFT (sTDDFT),\cite{Gao2015} optimally-tuned range separated hybrid
DFT~\cite{Neuhauser2015} and Bethe-Salpeter equation (sBSE).\cite{Rabani2015}
Among these, the closest to this work are the stochastic version of
sMP2 in real-time plane-waves,\cite{Neuhauser2013,Ge2013} and MO-based
MP2 with Gaussian basis sets.\cite{Ge2013} The stochastic method
presented here benefits from the fact that the GF2 self-energy is
a smooth function of imaginary time and is therefore naturally amenable
to random sampling. 

\section{Method}

\subsection{Brief review of GF2}

Our starting point is a basis of $N$ real single-electron non-orthogonal
atomic-orbital (AO) states $\phi_{i}\left(\mathbf{r}\right)$, with
an $N\times N$ overlap matrix $S_{ij}=\left\langle \phi_{i}\left|\phi_{j}\right.\right\rangle $.
Such states could be of any form, Gaussian, numerical, etc., but for
efficiency should be localized. We then use second quantization creation
$a_{i}^{\dagger}$ and annihilation $a_{i}$ operators with respect
to the non-orthogonal basis $\phi_{i}\left(\mathbf{r}\right)$. The
non-orthogonality is manifested only in a modified commutation relation, 

\begin{equation}
\{a_{i},a_{j}^{+}\}=(S^{-1})_{ij}.
\end{equation}
The Hamiltonian for the interacting electrons has the usual form
\begin{equation}
\hat{H}=\sum_{ij}h_{ij}a_{i}^{\dagger}a_{j}+\frac{1}{2}\sum_{ijkl}v_{ijkl}a_{i}^{\dagger}a_{k}^{\dagger}a_{l}a_{j},
\end{equation}
where $h_{ij}=\int d\boldsymbol{r}\phi_{j}\left(\boldsymbol{r}\right)\left(-\frac{1}{2}\nabla^{2}+v_{ext}\left(\boldsymbol{r}\right)\right)\phi_{i}\left(\boldsymbol{r}\right)$
and $v_{ext}\left(\boldsymbol{r}\right)$ is the bare external potential
(due to the nuclei), while $\hat{V}$ is the two electron-electron
(e-e) Coulomb interaction described by the 2-electron integrals 
\begin{equation}
v_{ijkl}=\iint\phi_{i}\left(\boldsymbol{r}\right)\phi_{j}\left(\boldsymbol{r}\right)v\left(\left|\boldsymbol{r}-\boldsymbol{r}^{\prime}\right|\right)\phi_{k}\left(\boldsymbol{r}^{\prime}\right)\phi_{l}\left(\boldsymbol{r}^{\prime}\right)d\boldsymbol{r}d\boldsymbol{r}^{\prime},\label{eq:2EI}
\end{equation}
where $v\left(r\right)=\frac{1}{r}$ is the Coulomb interaction potential. 

At a finite temperature $\beta^{-1}$ and chemical potential $\mu$
we employ the grand canonical density operator $\frac{e^{-\beta(\hat{H}-\mu\hat{N})}}{Z}$,
where $\hat{N}=\sum_{ij}S_{ij}a_{i}^{\dagger}a_{i}$ is the electron-number
operator and $Z\left(\beta\right)=\mbox{Tr}\left[e^{-\beta(\hat{H}-\mu\hat{N})}\right]$
is the partition function. The thermal expectation value of any operator
$\hat{A}$ can be calculated as $\left\langle \hat{A}\right\rangle =\mbox{Tr}\left[\frac{e^{-\beta((\hat{H}-\mu\hat{N})}}{Z\left(\beta\right)}\hat{A}\right].$
For one-body observables $\hat{A}=\sum_{ij}A_{ij}a_{i}^{\dagger}a_{j}$
we write $\left\langle \hat{A}\right\rangle =\sum_{ij}A_{ij}P_{ij}$
where $P_{ij}=\left\langle a_{i}^{\dagger}a_{j}\right\rangle $ is
the reduced density matrix.

The 1-particle Green's function $G_{jk}\left(\tau\right)$ at an imaginary
time $\tau$ is a generalization of the concept of the density matrix
and obeys an equation of motion that can be solved by perturbation
methods. Formally:
\begin{equation}
G_{jk}\left(\tau\right)=-\left\langle Ta_{j}\left(\tau\right)a_{k}^{\dagger}\right\rangle ,\label{eq:def-G(tau)}
\end{equation}
where $a_{j}\left(\tau\right)\equiv e^{(\hat{H}-\mu\hat{N})\tau}a_{j}e^{-(\hat{H}-\mu\hat{N})\tau}$
with $-\beta<\tau<\beta$, and $T$ is the time-ordering symbol: 
\begin{equation}
Ta_{j}\left(\tau\right)a_{k}^{\dagger}\equiv\theta\left(\tau\right)a_{j}\left(\tau\right)a_{k}^{\dagger}-\theta\left(-\tau\right)a_{k}^{\dagger}a_{j}\left(\tau\right).
\end{equation}
Note that $G(\tau)$ is a real and symmetric matrix.

Each element $G_{jk}\left(\tau\right)$ (and therefore the entire
matrix $G\left(\tau\right)$) is discontinuous when going from negative
to positive times, but this discontinuity is not a problem since we
only need to treat explicitly positive times $\tau>0$ while negative
$\tau$'s are accessible by the anti-periodic relation for $G\left(\tau\right)$
\begin{equation}
G\left(\tau\right)=-G\left(\tau+\beta\right),\,\,\,-\beta<\tau<0,
\end{equation}
as directly verified by substitution in Eq.~(\ref{eq:def-G(tau)}).
Hence $G\left(\tau\right)$ can be expanded as a Fourier series involving
the Matsubara frequencies $\omega_{n}=\left(2n+1\right)\frac{\pi}{\beta}$
:
\begin{align}
G\left(\tau\right) & =\frac{1}{\beta}\sum_{n=-\infty}^{\infty}G\left(i\omega_{n}\right)e^{-i\omega_{n}\tau}\label{eq:G(tau)-matsubara}
\end{align}
where: 
\begin{equation}
G\left(i\omega_{n}\right)=\int_{0}^{\beta}G\left(\tau\right)e^{i\omega_{n}\tau}d\tau.\label{eq:G(iwn)-matsubara}
\end{equation}

The Green's function of Eq.~(\ref{eq:def-G(tau)}) gives access to
the reduced density matrix by taking the imaginary time $\tau$ as
a negative infinitesimal (denoted as $0^{-})$:dxn@ucla.edu
\begin{align}
P_{kj} & =2G_{kj}\left(0^{-}\right)=-2G_{kj}(\beta^{-})\label{eq:DM}\\
 & =\frac{2}{\beta}\sum_{n=0}^{\infty}e^{-i\omega_{n}0^{-}}G_{kj}\left(i\omega_{n}\right).\nonumber 
\end{align}
 Hence, all thermal averages of one-electron operators are accessible
through the sum of the Matsubara coefficients. 

Perturbation theory can be used to build approximations for $G\left(\tau\right)$
based on a non-interacting Green's function $G_{0}\left(\tau\right)$
corresponding to a reference one-body Hamiltonian $\hat{H_{0}}=\sum_{ij}F_{ij}a_{i}^{\dagger}a_{j}$.
Here, $F$ is any real symmetric ``Fock'' matrix such that $\hat{H_{0}}$
well approximates the interacting electron Hamiltonian. The derivation
of $G_{0}\left(\tau\right)$ requires orthogonal combination of the
basis set, i.e., finding a matrix $X$ that fulfills $XX^{T}=S^{-1}$.
Then it is straightforward to show that 

\begin{align}
G_{0}\left(\tau\right) & =\label{eq:NI-G}\\
 & Xe^{-\tau\left(\bar{F}-\mu\right)}\left[\frac{\theta\left(-\tau\right)}{1+e^{\beta\left(\bar{F}-\mu\right)}}-\frac{\theta\left(\tau\right)}{1+e^{-\beta\left(\bar{F}-\mu\right)}}\right]X^{T}\nonumber 
\end{align}
where $\bar{F}=X^{T}FX$ is the Fock matrix in the orthogonal basis
set. Note that for positive (or negative) imaginary times $G_{0}\left(\tau\right)$
is a real, smooth and non-oscillatory Green's function. This is important
for us since it much easier to stochastically sample a smooth function. 

\begin{figure}
\includegraphics[scale=0.25]{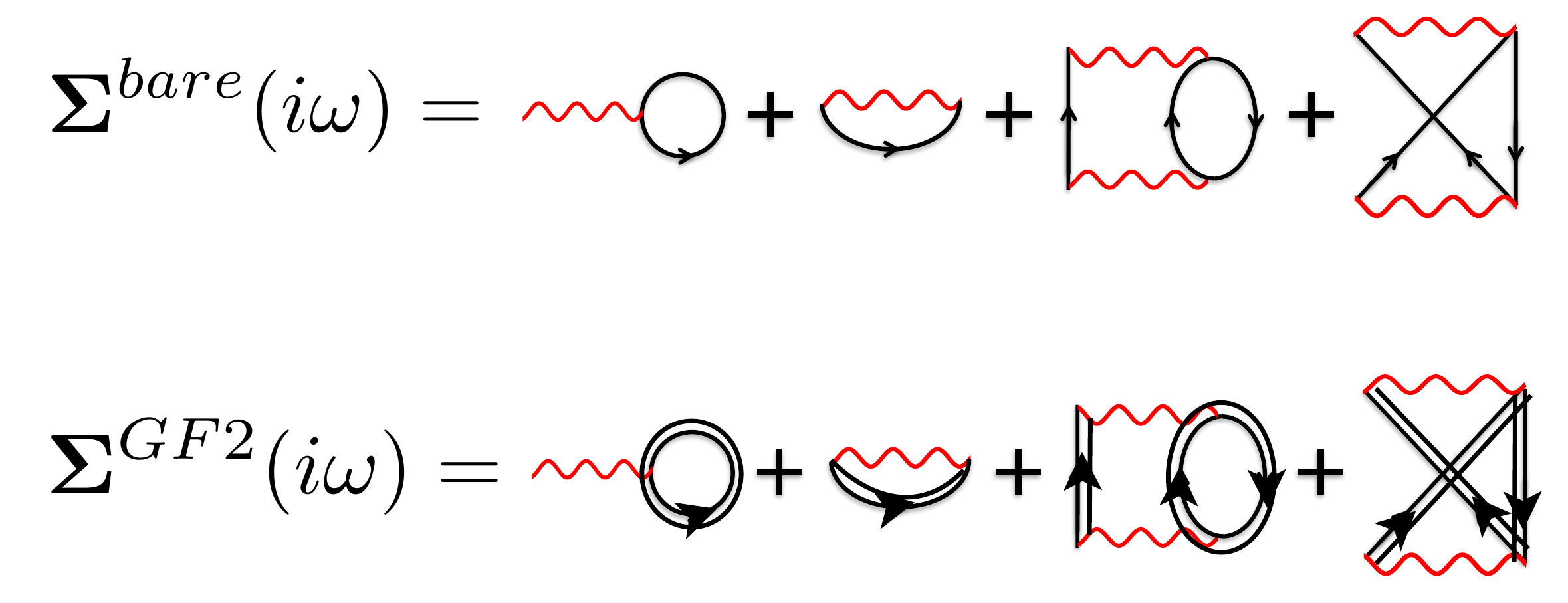} \caption{\label{Fig:SE_GF2}Upper panel: Bare second order self-energy diagrams.
Lower panel: Second order self-energy diagrams evaluated self-consistently.
Note that the Green's function lines are renormalized while the interactions
lines remain bare. For details see Ref.~\citenum{Phillips2014}.}
\end{figure}

Integration of Eqs.~(\ref{eq:NI-G}) yields: 
\begin{equation}
G_{0}\left(i\omega_{n}\right)=\left((\mu+i\omega_{n})S-F\right)^{-1}.\label{eq:G0w}
\end{equation}
Since we now know how to write down Green's functions for non-interacting
systems, we rewrite the unknown part of the exact Green's function
by introducing the frequency-dependent self-energy, formally defined
by:
\begin{equation}
G\left(i\omega_{n}\right)=\left(\left(\mu+i\omega_{n}\right)S-F-\Sigma\left(i\omega_{n}\right)\right)^{-1},\label{eq:Basic-G}
\end{equation}
and by construction the self-energy fulfills the Dyson equation:
\begin{equation}
G\left(i\omega_{n}\right)=G_{0}\left(i\omega_{n}\right)+G_{0}\left(i\omega_{n}\right)\Sigma\left(i\omega_{n}\right)G\left(i\omega_{n}\right).\label{eq:Dyson}
\end{equation}
Instead of viewing these equations as a definition of the self-energy
$\Sigma\left(i\omega_{n}\right)$, we can calculate this self-energy
to a given order of perturbation theory in $\Delta\hat{H}=\hat{H}-\hat{H}_{0}$.
Specifically, the GF2 approximation\cite{Phillips2014,Dahlen05} uses
a Hartree-Fock ansatz for $F$, 
\begin{align}
F_{ij} & =h_{ij}+\frac{1}{2}P_{kl}\left(2v_{ijkl}-v_{ilkj}\right),\label{eq:FockMat}
\end{align}
where an Einstein summation convention is used, summing indices that
appear in pairs (here, both $k$ and $l$). The self-energy in imaginary
time $\Sigma(\tau)$ is then obtained by second order perturbation
theory (see Fig.~(\ref{Fig:SE_GF2})): 
\begin{align}
\Sigma_{ij}\left(\tau\right) & =G_{kl}\left(\tau\right)G_{mn}\left(\tau\right)G_{pq}\left(\beta-\tau\right)v_{impk}\left(2v_{jnlq}-v_{jlnq}\right).\label{eq:sigma(tau)}
\end{align}
Note that $\Sigma(\tau)$ and $\Sigma(i\omega_{n})$ are connected
by exactly the same Matsubara relations connecting $G(\tau)$ and
$G(i\omega_{n}),$ Eqs. \ref{eq:G(tau)-matsubara}-\ref{eq:G(iwn)-matsubara}.

The self-consistent one-body Green's function governs all one-body
expectation values. Moreover, even the total two-body potential energy
is available, by differentiation of the matrix trace (denoted by $\mbox{Tr}\left[\right]$)
of the Green's function with respect to $\tau$: $\left\langle \hat{V}\right\rangle =-\frac{1}{2}\lim_{\tau\to0^{-}}\mbox{Tr}\left[\left((\frac{\partial}{\partial\tau}-\mu)S+h\right)G\left(\tau\right)\right]$.
Hence, the total energy is:
\begin{equation}
\left\langle \hat{H}\right\rangle ={\rm Tr}\left[hP-\frac{1}{2}\lim_{\tau\to0^{-}}\left((\frac{\partial}{\partial\tau}-\mu)S+h\right)G\left(\tau\right)\right].\label{eq:Etot_tauderivative}
\end{equation}
It is easy to show by plugging the definition of $G(\tau)$ to Eq.
(\ref{eq:Etot_tauderivative}) that this total energy has convenient
frequency and time forms:

\begin{align}
\left\langle \hat{H}\right\rangle  & =\frac{1}{2}\mbox{Tr}\left[\left(h+F\right)P\right]+\frac{2}{\beta}\mbox{Re}\sum_{n}\mbox{Tr}\left[G\left(i\omega_{n}\right)\Sigma^{T}\left(i\omega_{n}\right)\right]\label{eq:Etot}\\
 & =\frac{1}{2}\mbox{Tr}\left[\left(h+F\right)P\right]+2\int_{0}^{\beta}\text{Tr}\left[G\left(\beta-\tau\right)\Sigma\left(\tau\right)\right]d\tau.\nonumber 
\end{align}

To conclude, the combination of Eqs. (\ref{eq:DM}), (\ref{eq:Basic-G}),
(\ref{eq:FockMat}) and (\ref{eq:sigma(tau)}) along with the requirement
that the density matrix describes $N_{e}$ electrons results in the
following self-consistent GF2 procedure:
\begin{enumerate}
\item Perform a standard HF calculation and obtain a starting guess for
the Fock matrix $F=F_{HF}$ and the density matrix $P=P_{HF}.$ Set
$\Sigma\left(i\omega_{n}\right)=0$ for the set of $N_{\omega}$ positive
Matsubara frequencies $\omega_{n},$ $n=0,1,2,\dots,N_{\omega}-1$.,
.
\item \label{enu:determine-miu}Given $\Sigma\left(i\omega_{n}\right)$
and $F$, find $\mu$ such that ${\rm Tr}\left[PS\right]=N_{e}$,
where $P$ is given in Eq.~(\ref{eq:DM}) from $G(\tau=\beta^{-})$
which depends on $\mu$ through the basic definition Eq.~(\ref{eq:Basic-G}).
\item \label{enu:Calculate-G(tau)}Calculate $G\left(\tau\right)$ (Eq.~(\ref{eq:G(tau)-matsubara}))
and $P$ (Eq.~(\ref{eq:DM})).
\item \label{enu:Calculate-F}Calculate the Fock matrix $F$ from $P$ (Eq.~(\ref{eq:FockMat})).
\item \label{enu:calc-Sigma}Calculate the self-energy $\Sigma\left(\tau\right)$
from Eq.~(\ref{eq:sigma(tau)}) and transform is to the Matsubara
frequency domain to yield $\Sigma\left(i\omega_{n}\right).$
\item Calculate the total energy$\left\langle \hat{H}\right\rangle $ from
Eq. (\ref{eq:Etot}).
\item Repeat steps 2-6 until convergence of the density and the total energy.
\end{enumerate}
Once converged, the GF2 correlation energy is defined as the difference
$E_{{\rm corr}}=\left\langle \hat{H}\right\rangle -E_{{\rm HF}}$
between the converged total energy (Eq.~(\ref{eq:Etot})) and the
initial Hartree-Fock energy, $E_{HF}=\frac{1}{2}\mbox{Tr}\left[\left(h+F_{HF}\right)P_{HF}\right]$.
Note that in the first iteration GF2 yields automatically the temperature-dependent
MP2 energy: 

\begin{align}
E_{{\rm MP2}}^{corr} & =\int_{0}^{\beta}\text{Tr}\left[G_{0}\left(\beta-\tau;F_{HF}\right)\Sigma_{0}\left(\tau\right)\right]d\tau,\label{eq:EMP2}
\end{align}
where $\Sigma_{0}\left(\tau\right)$ is that of Eq.~(\ref{eq:sigma(tau)})
with $G_{0}$ replacing $G$. This expression reduces to the familiar
MP2 energy expression at the limit $\beta\to\infty$ (zero temperature
limit), when evaluated in the molecular orbital basis set that diagonalizes
the matrix $F_{HF}$. 

Finally, a technical point. The representation of the Green's functions
in $\tau$-space can be complicated when the energy range of the eigenvalues
of $F$ is large since a function of the type $e^{-\tau\left(f-\mu\right)}/\left(1+e^{-\beta\left(f-\mu\right)}\right)$
can be spiky when $f>\mu$ and $\tau\to\beta$ or when $f<\mu$ and
$\tau\to0$. This requires special techniques for both imaginary time
and frequency grids as discussed in Refs.~\citenum{Kananenka2016,Kananenka16}.

\subsection{sGF2: Stochastic approach to GF2}

Most of the computational steps in the above algorithm scale with
system size $N$ (number of AO basis functions) as ${\rm O}\!\left(N_{SC}\times N_{\tau}\times N^{3}\right)$
where $N_{SC}$ is the number of GF2 self-consistent iterations and
$N_{\tau}$ is the number of time-steps. However, the main numerical
challenge in GF2 is step \ref{enu:calc-Sigma} (Eq.~(\ref{eq:sigma(tau)}))
which scales formally as ${\rm O}\!\left(N_{SC}\times N_{\tau}\times N^{5}\right)$
making GF2 highly expensive for any reasonably sized system. This
steep scaling is due to the contraction of two 4-index tensors with
three Green's function matrices. 

To reduce this high complexity, we turn to the stochastic paradigm
which represents the matrices $G\left(\tau\right)$ by an equivalent
random average over stochastically chosen vectors. Fundamentally,
this is based on resolving the identity operator. Specifically, for
each $\tau$ we generate a vector $\eta^{0}$ of $N$ components randomly
set to $+1$ or $-1$. Vectors at different times $\tau$ are statistically
independent, but we omit for simplicity their $\tau$ labeling. Then,
the key, and trivial, observation is that average of the product of
different components of $\eta^{0}$ is the unit matrix, which we write
symbolically as 

\begin{equation}
\eta_{k}^{0}\eta_{l}^{0}=\delta_{kl}.\label{eq:random-delta}
\end{equation}
We emphasize that the equality in this equation should be interpreted
to hold in the limit of averaging over infinitely many random vectors
$\eta^{0}.$ 

Given this separable presentation of the unit matrix, it is easy to
rewrite any matrix as an average over separable vectors. Specifically,
from $\eta^{0}$ we define the two vectors: 

\begin{equation}
\eta=\sqrt{|G\left(\tau\right)|}\eta^{0},\,\,\,\,\,\bar{\eta}=sgn\left(G\left(\tau\right)\right)\sqrt{|G\left(\tau\right)|}\eta^{0},\label{eq:eta_sqrtG_etabar_sign}
\end{equation}
and then

\begin{equation}
G_{kl}\left(\tau\right)=\bar{\eta}_{k}\eta_{l}.\label{eq:Gtau-rand}
\end{equation}
Here, the square-root matrix is $\sqrt{\left|G\left(\tau\right)\right|}=A\sqrt{\left|g\right|}A^{T}$,
where $A\left(\tau\right)$ is the unitary matrix of eigenvectors
and $g\left(\tau\right)$ is the diagonal matrix of eigenvalues of
$G\left(\tau\right)$. 

As a side note, we have a freedom to choose other vectors; specifically,
any two vectors $\bar{\eta}=\bar{D}\eta^{0},\,\,\,\,\,\eta=D\eta^{0}$,
will work if $\bar{D}D^{T}=G(\tau).$ In principle, we can even use
the simplest choice $\bar{D}=1,\,\,\,\,\,D=G(\tau),$ corresponding
to $\bar{\eta}=\eta^{0}$ and $\eta=G(\tau)\eta^{0}$. But while this
latter choice has the advantage that $G(\tau)$ does not need to be
diagonalized, we find that it is numerically better to use Eq. (\ref{eq:eta_sqrtG_etabar_sign})
as it is more balanced and therefore converges faster with the number
of stochastic samples. Also note that at the first iteration, where
$G(\tau)=G_{0}(\tau),$ there is no need to diagonalize $G_{0}(\tau)$
at different times, since it is obtained directly from the eigenstates
of $\bar{F}$ in Eq. (\ref{eq:NI-G}).

Going back to Eq. (\ref{eq:eta_sqrtG_etabar_sign}), we similarly
separate the other two Green's function matrices appearing in Eq.~(\ref{eq:sigma(tau)}),
writing them as $G_{mn}\left(\tau\right)=\bar{\xi}_{m}\xi_{n}$ and
$G_{pq}\left(\beta-\tau\right)=\bar{\zeta}_{p}\zeta_{q}$. The self-energy
in Eq.~(\ref{eq:sigma(tau)}) is then
\begin{align}
\Sigma_{ij}\left(\tau\right) & =\bar{\eta}_{k}\bar{\xi}_{m}\bar{\zeta}_{p}v_{impk}\left(2\eta_{l}\xi_{n}\zeta_{q}v_{jnlq}-\eta_{l}\xi_{n}\zeta_{q}v_{jlnq}\right),
\end{align}
so that it is separable to a product of two terms
\begin{equation}
\Sigma_{ij}\left(\tau\right)=\bar{u}_{i}\left[2u_{j}-w_{j}\right],\label{eq:StochSelfEn}
\end{equation}
where we defined three auxiliary vectors 
\begin{align}
\bar{u}_{i} & =\bar{\eta}_{k}\bar{\xi}_{m}\bar{\zeta}_{p}v_{impk}\nonumber \\
u_{j} & =\eta_{l}\xi_{n}\zeta_{q}v_{jnlq}\label{eq:uiubjwi}\\
w_{i} & =\eta_{l}\xi_{n}\zeta_{q}v_{jlnq}.\nonumber 
\end{align}
The self-energy in Eq. (\ref{eq:StochSelfEn}) should be viewed as
the average, over the stochastic vectors $\xi^{0}$, $\eta^{0}$ and
$\zeta^{0},$ of the product term ($\bar{u}_{i}$ times $2u_{j}-w_{j}).$

The direct calculation of the vectors $\bar{u},\,u,\,w$ by Eq. (\ref{eq:uiubjwi})
is numerically expensive once $M>30$ . We reduce the scaling by recalling
the definition of $v_{jnlq}$ in Eq.~(\ref{eq:2EI}):
\begin{align}
u_{j} & =\eta_{l}\xi_{n}\zeta_{q}\iint\phi_{j}\left(\boldsymbol{r}\right)\phi_{n}\left(\boldsymbol{r}\right)v\left(\left|\boldsymbol{r}-\boldsymbol{r}^{\prime}\right|\right)\phi_{l}\left(\boldsymbol{r}^{\prime}\right)\phi_{q}\left(\boldsymbol{r}^{\prime}\right)d\boldsymbol{r}d\boldsymbol{r}^{\prime}\\
 & =\iint\phi_{j}\left(\boldsymbol{r}\right)\xi\left(\boldsymbol{r}\right)v\left(\left|\boldsymbol{r}-\boldsymbol{r}^{\prime}\right|\right)\eta\left(\boldsymbol{r}^{\prime}\right)\zeta\left(\boldsymbol{r}^{\prime}\right)d\boldsymbol{r}d\boldsymbol{r}^{\prime},
\end{align}
where:
\begin{equation}
\eta\left(\boldsymbol{r}\right)=\eta_{l}\phi_{l}\left(\boldsymbol{r}\right),\label{eq:eta(R)}
\end{equation}
and $\xi\left(\boldsymbol{r}\right)$ and $\zeta\left(\boldsymbol{r}\right)$
are analogously defined. We can therefore write

\begin{equation}
u_{j}=\int\phi_{j}\left(\boldsymbol{r}\right)\xi\left(\boldsymbol{r}\right)v_{\eta\zeta}\left(\boldsymbol{r}\right)d\boldsymbol{r},\label{eq:uj}
\end{equation}
where
\begin{equation}
v_{\eta\zeta}\left(\boldsymbol{r}\right)\equiv\int v\left(\left|\boldsymbol{r}-\boldsymbol{r}^{\prime}\right|\right)\eta\left(\boldsymbol{r}^{\prime}\right)\zeta\left(\boldsymbol{r}^{\prime}\right)d\boldsymbol{r}^{\prime}\label{eq:vHartree}
\end{equation}
is the Coulomb potential corresponding to the random charge distribution
$\eta\left(\boldsymbol{r}\right)\zeta\left(\boldsymbol{r}\right)$.
Similar expressions apply for $\bar{u}_{i}$ and $w_{j}$. 

Equations (\ref{eq:uj})-(\ref{eq:vHartree}) are performed numerically
using FFT methods on a 3D Cartesian grid with $N_{g}$ grid points,
so Eq.~(\ref{eq:vHartree}) is calculated with ${\rm O}\!\left(N_{g}\log N_{g}\right)$
operations. Since the AO basis functions $\phi_{i}\left(\boldsymbol{r}\right)$
are local in 3D space, the calculations of $\eta\left(\boldsymbol{r}\right)$,
$\xi\left(\boldsymbol{r}\right)$ and $\zeta\left(\boldsymbol{r}\right)$
in Eq.~(\ref{eq:eta(R)}) scale linearly with system size. 

Eq.~(\ref{eq:StochSelfEn}) gives an exact expression for $\Sigma_{ij}\left(\tau\right),$
as an expected value over formally an infinite number of stochastic
orbitals $\eta^{0}$, $\xi^{0}$ and $\zeta^{0}$. Actual calculations
use a finite number $I$ of ``stochastic iterations'', where in
each such iteration a set of stochastic vectors $\eta^{0}$, $\xi^{0}$
and $\zeta^{0}$ (different at each $\tau)$ is generated and $\Sigma_{ij}\left(\tau\right)$
is averaged over them. The overall scaling of this step is therefore
$I\times N_{\tau}\times\left(N_{g}\log N_{g}+N^{2}\right)$. We note
that the typical values of $N_{\tau}$ and $I$ are in the hundreds,
see the discussion of the stochastic error below. 

Finally, we note that while the stochastic vectors ($\eta^{0}$,$\zeta^{0},\xi^{0}$)
are statistically independent for each time point $\tau$, the same
$\tau-$dependent vectors are used at each GF2 iteration, making it
possible to converge these iterations. 

\begin{figure*}
\includegraphics[width=1\textwidth]{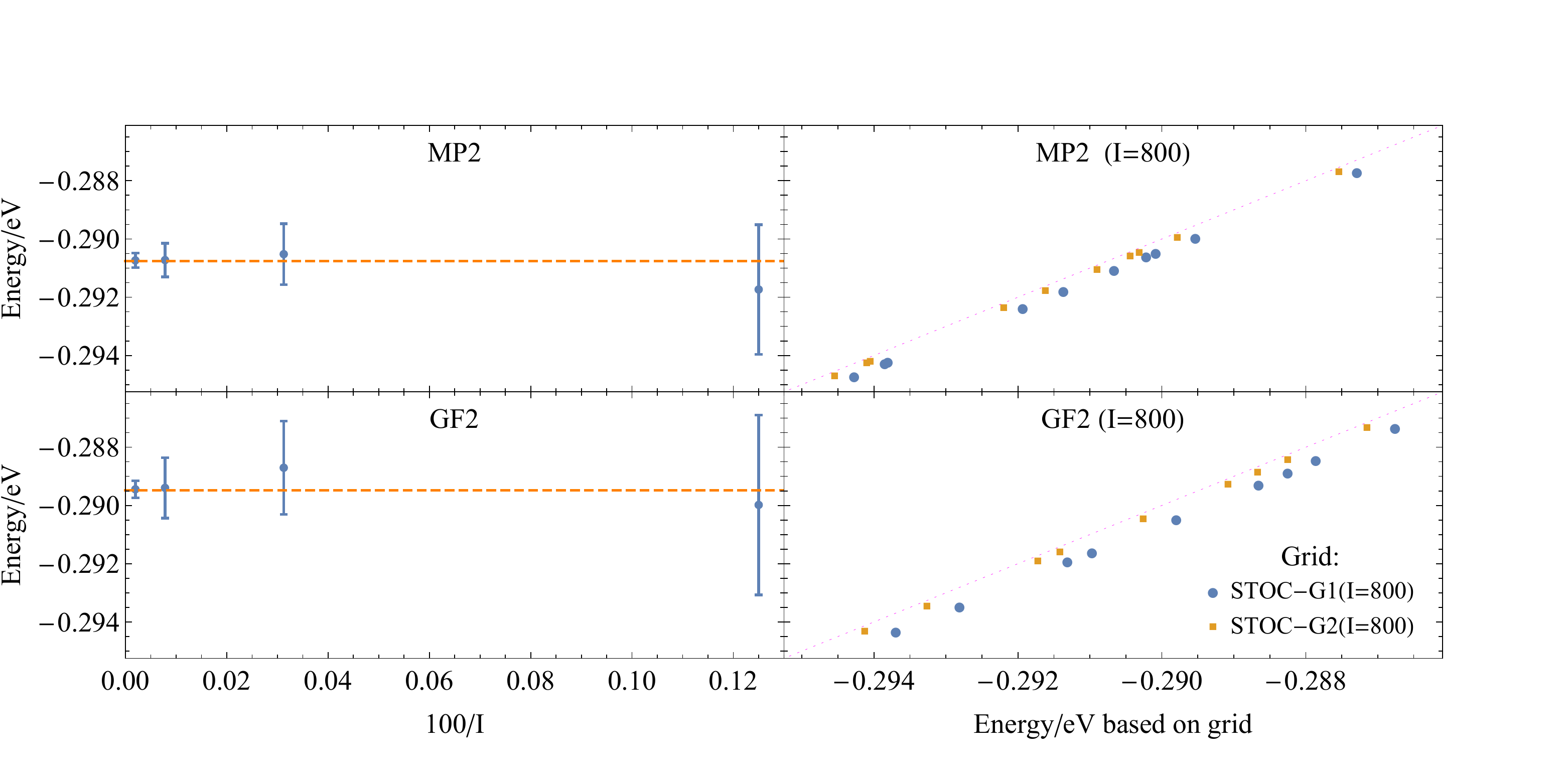}

\caption{\label{fig:Validation-H10}MP2 and GF2 correlation energies per electron
for a linear $\text{H}_{10}$ system with nearest neighbor spacing
of 1�. Left panels: the DET (deterministic) correlation energies (dashed
horizontal lines) are well within the $I$-dependent error-bars $\bar{E}\pm2\sigma$
of the STOC-NG calculations, where $\bar{E}$ and $\sigma$ are the
average and standard deviation of the correlation energy calculated
in 10 statistically independent runs. Right panels: a correlation
plot of pairs of stochastically ($I=800$) estimated correlation energies:
$\left(E_{STOC-G1\left(800\right)},E_{STOC-NG\left(800\right)}\right)_{i}$
as blue dots and $\left(E_{STOC-G2\left(800\right)},E_{STOC-NG\left(800\right)}\right)_{i}$
as orange dots, $i=1,\dots,10$. The diagonal dotted line represents
the perfect correlation $E_{STOC-G1\,or\,2}=E_{STOC-NG}$. }
\end{figure*}

\section{Results}

\subsection{Systems and specifics}

The algorithm was tested on linear hydrogen chains, $(H_{M}$) a nearest
neighbor distance of $1$�, for several sizes: $M=10,100,300$ and
$1000$. The linearity was for convenience and we emphasize that it
does not play any role in the algorithm. The smallest chain was used
to demonstrate the convergence of the approach to the basis-set deterministic
values, and the other three calculations were used to study the dependence
of the algorithm on system size. 

In all calculations, an STO-3G basis was used, so that in this case
$N=M$ and obviously the number of electrons is also $N_{e}=M$. A
periodic spatial grid of $0.5a_{0}$ spacing was used to represent
the wave functions, and the grids contained $10\times10$ points in
the direction orthogonal to chain and between 60 and 4000 points along
the chain, depending on system size. For the smallest system ($H_{10}$)
a finer, bigger grid was also used, as detailed below. 

Other, technical details: 
\begin{itemize}
\item Periodic images were screened using the method of Ref.~\citenum{Martyna1999}.
\item The inverse temperature was $\beta=50E_{h}^{-1}$. 
\item A Chebyshev-type imaginary-time grid with 128 time points was employed
using a spline-fit method \cite{Kananenka2016,Kananenka16} for the
frequency-to-time conversions of $G(i\omega_{n})$ and $\Sigma(i\omega_{n})$
and for the evaluation of the two-body energy.
\end{itemize}

\subsection{Small system}

In our GF2 and MP2 algorithm, we make two types of numerical discretizations.
First, we use a finite number (labeled $I$) of stochastic iterations
to sample the self-energy, so we must show convergence as $I$ grows.
Second, we use grids for bypassing the need to sum over ${\rm O\!}\left(N^{4}\right)$
two-electron integrals, hence we need to demonstrate convergence with
respect to grid quality. We therefore examine in this section a small
system, linear $\text{H}_{10}$, and make four types of GF2/MP2 correlation
energy calculations: 
\begin{itemize}
\item DET: fully deterministic calculations based on the analytical 2-electron
integrals; 
\item STOC($I$)-NG: stochastic calculations based on $I$ stochastic iterations
and on the analytical two electron integrals; 
\item STOC($I$)-G1 and STOC($I$)-G2: stochastic calculations based on
$I$ stochastic iterations and on a 3D grid. Here, G1 is the same
type of grid we use for the larger calculations, and includes $10\times10\times60$
points with a spacing $h=0.5a_{0}$. G2 is somewhat denser and covers
more space, with $16\times16\times100$ points and $h=0.4a_{0}$. 
\end{itemize}
Our strategy is to first show that STOC-NG($I$) converges to the
deterministic set (DET) as $I$ grows. Then we show that for a given
number of stochastic orbitals, $I=800,$ both grid results are quite
close to the non-grid result, and that the somewhat better second
grid (STOC($I=800$)-G2 leads to extremely close results to the non-grid
values (STOC($I=800$)-NG), so that the convergence with grid is very
rapid. 

We repeat the STOC-NG calculation 10 times determining the average
correlation energy $\bar{E}$ and its standard deviation $\sigma$
as a function of $I$. The results are shown in the left panels of
Fig.~\ref{fig:Validation-H10} as error-bars at $\bar{E}\pm\sigma$,
which shrink approximately as $1/\sqrt{I}$ and which include the
DET result, represented as dashed horizontal lines, showing very small
or no bias. For MP2, a bias in the stochastic calculations is not
expected since the correlation energy is calculated linearly from
the first iteration of the self-energy $\Sigma_{0}$ (Eq.~(\ref{eq:EMP2})).
But for GF2 such a bias may form since the the ``noisy'' self-energy
is used non-linearly to update the Green's function in Eq.~(\ref{eq:Basic-G}).
However, for this small $N=10$ system the stochastic MP2 and GF2
energies do not exhibit a noticeable bias. We discuss the bias in
larger systems below. 

Next, we asses the errors associated with using grid calculations
replacing the analytical 2-electron integration. In both right panels
of Fig.~\ref{fig:Validation-H10} we show 10 blue dots, each corresponding
to a pair of stochastic energies $\left(E_{STOC-G1\left(800\right)},E_{STOC-NG\left(800\right)}\right)_{i}$
, $i=1,\dots,10$, both calculated with the same random seed $s_{i}$
(of course $s_{i}$ and $s_{j}$ are statistically independent). We
also show 10 orange dots, each corresponding to a pair of stochastic
energies $\left(E_{STOC-G2\left(800\right)},E_{STOC-NG\left(800\right)}\right)_{i}$,
also calculated with the same seed $s_{i}$ as before. The use of
the same seeds for each pair of blue and orange dots allows for comparison
of the grid error (which is the horizontal distance of a point from
the diagonal) without worrying about the larger statistical error,
seen as the spread of the results along the diagonal. We see that
the grid error decreases significantly when moving from G1 to G2,
but even the error for G1 is already very small (about 0.5meV per
electron). 

\subsection{Larger systems}

\begin{figure}
\includegraphics[width=1\columnwidth]{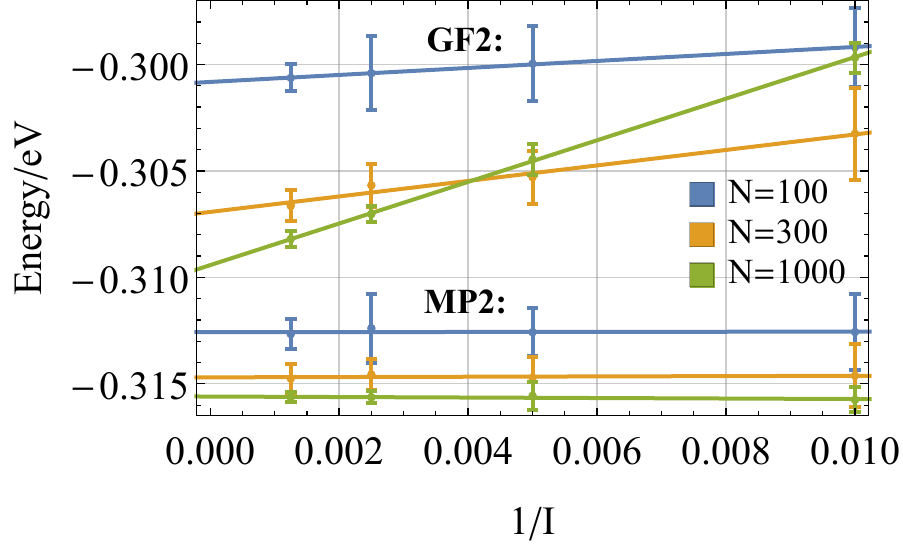}

\caption{\label{fig:Statistics}Statistical estimates of the MP2 and GF2 correlation
energies per electron for three linear chains with $N=100$, $300$
and $1000$ H atoms as described in the text, as a function of the
inverse number of sets of stochastic orbitals $I$. The results are
shown as error bars, where the center of each error bar is the average
and its width is the standard deviation of the correlation energy
estimates from 10 statistically independent runs, each employing $I$
sets of stochastic orbitals. The lines are linear regression fits
to the data and uncertainties.}
\end{figure}

In the small system considered above the bias was not noticeable and
here we examine the bias in larger systems. In Fig.~\ref{fig:Statistics}
we show the STOC-G1(I) correlation energies in three specific systems
composed of $N=100,300$ and $1000$ hydrogen atoms placed on a straight
line with a nearest neighbor spacing of $1$ �. 

We first study the MP2 correlation energy of each system, appearing
in the lower energy range in the figure. The starting point of the
GF2 calculation is the Hartree-Fock $F_{HF}$ and $P_{HF}$ matrices,
so the MP2 energy is half the correlation energy of the first self-consistent
iteration (see Eqs. (\ref{eq:Etot}) and (\ref{eq:EMP2})). The statistical
errors in MP2 are pure fluctuations, a random number distributed normally
with zero average and with standard deviation given by $\sigma_{0}/\sqrt{I}$
where $\sigma_{0}$ is independent of $I$ but shrinks with chain
length: $\sigma_{0}\propto1/\sqrt{L}$, exhibiting ``self averaging''.~\cite{Baer2013}
The stochastic MP2 errors are very small and decrease with system
size, so for $N=1000$ the standard deviation of the $I=800$ iteration
calculation is 0.07\% of the total correlation energy. For perspective,
note that (deterministic) errors of larger or similar magnitude are
present in linear scaling local or divide and conquer MP2 methods
with density fitting.~\cite{Werner2003,Baudin2016}

Next, we discuss the stochastic estimates of the self-consistent GF2
correlation energies. These exhibit statistical errors with two visible
components. The first is a \emph{fluctuation}, similar in nature to
that of the MP2 calculation, and the second component is a \emph{bias
}which decreases as $I$ grows. In fact, we expect the bias to asymptotically
decrease inversely with $I$, \footnote{A bias arises whenever we plug a random variable $x$, having an expected
value $\mu$ and variance $\sigma^{2}$, into a nonlinear function
$f\left(x\right)$. One cannot hope that $f\left(x\right)$ will have
the expected value of $f\left(\mu\right)$ unless $f$ is a linear
function. A simple example is $f\left(x\right)=x^{2}$, where from
the definition of variance \unexpanded{$\langle f(x) \rangle = f(\mu)+\sigma^2$}
. Using the Taylor expansion of $f$ around $\mu$, it is straightforward
to show that $f\left(\bar{x}\right)$, where \unexpanded{$\bar{x}=\frac{1}{I}\sum_{i=1}^{I}x_i$}
is an average over $I$ samples and when $I$ is sufficiently large,
\unexpanded{$\langle f(\bar{x})\rangle\approx f(\mu)+\frac{f''(\bar{x})\sigma^2}{2I}$}
and so the bias is proportional to the variance of $x$, the curvature
of $f$ at $\mu$ and inversely proportional to the number of iterations
$I$. } so we fit the numerical GF2 results to a straight line in $I^{-1}$.
Table~\ref{tab:The-bias} shows the estimate of the correlation energies,
the fluctuation and the bias as a function of the number of stochastic
orbitals. The results are highly accurate, for example when $I=800$
is used for the largest system ($N=1000$), the errors in MP2 and
in GF2 are smaller than 0.1\%.

\begin{table*}
\begin{tabular}{|c|c|c||c|c|}
\hline 
$N$ & \multicolumn{2}{c||}{MP2 Energy (eV)} & \multicolumn{2}{c|}{GF2 Energy (eV)}\tabularnewline
\hline 
 & $E\left(N,I\right)$ & $I=800$ & $E\left(N,I\right)$ & $I=800$\tabularnewline
\hline 
\hline 
100 & $-0.3126\pm0.025I^{-1/2}$  & $-0.3127(9)$  & $-0.3008+0.16I^{-1}\pm0.025I^{-1/2}$  & $-0.3007(9)$ \tabularnewline
\hline 
300 & $-0.3148\pm0.015I^{-1/2}$ & $-0.3148(5)$ & $-0.3069+0.36I^{-1}\pm0.020I^{-1/2}$ & $-0.3067(7)$\tabularnewline
\hline 
1000 & $-0.3157\pm0.007I^{-1/2}$ & $-0.3157(2)$ & $-0.3094+0.98I^{-1}\pm0.010I^{-1/2}$ & $-0.3094(3)$\tabularnewline
\hline 
\end{tabular}

\caption{\label{tab:The-bias}The statistical estimate of the MP2 and GF2 energies
per electron of the linear hydrogen chains $\text{H}_{N}$, based
on runs with up to $I=800$ stochastic samplings. For MP2 there is
no bias and the error is a only statistical fluctuation which for
$I=800$ is very small, on the order of 0.1\% of the correlation energy
or better. For GF2 the statistical fluctuation is similarly tiny,
and there is a bias (depending to leading order on $I^{-1})$ that
grows with $N$, and therefore $I$ should be similar to $N$ or larger.}
\end{table*}

{\em Timings.} The measured overall CPU time for the stochastic
self-energy calculation (performed on a XEON system) can be expressed
as
\begin{equation}
T^{\Sigma}\approx2.5\times N\times N_{\tau}\times I\times10^{-7}hr,
\end{equation}
where, as mentioned, $N$ is the number of electrons and $I$ the
number of stochastic orbitals in the system. The MP2 wall time calculation
is essentially equal to the self-energy time divided by the number
of cores $n_{CORES}$, since the parallelization has negligible overhead:
\begin{equation}
T_{\text{wall}}^{MP2}\approx\frac{T^{\Sigma}}{n_{CORES}}.
\end{equation}

GF2 involves an additional step, where the Green's function is constructed
from the self-energy and this step scales cubically with system size.
Furthermore, there are $N_{SC}$ self-consistent iterations. The total
time is therefore found to be:
\begin{align*}
T_{\text{wall}}^{GF2} & \approx N_{SC}\frac{\left(1.7\times N^{3}\times N_{\tau}\text{\ensuremath{\times}1\ensuremath{0^{-11}}hr}+T^{\Sigma}\right)}{n_{CORES}}.
\end{align*}
For the $\text{H}_{1000}$ system, with $I=800$ stochastic orbitals,
the MP2 calculation takes $T_{\text{wall}}^{MP2}=24\text{hr}/n_{CORES}$,
i.e. about $30$min when using 48 cores. 

The GF2 calculation for this same system involves $N_{SC}=12$ iterations
and a cubic part which takes about 2 core-hours per iteration, i.e.,
the cubic part is still an order of magnitude smaller than the self-energy
sampling time for this system size. The wall time is therefore $T_{\text{wall}}^{GF2}=6.5\text{hr}$
with 48 cores. 

For the $\text{H}_{300}$ system we find $T_{\text{wall}}^{MP2}=12$min
and $T_{\text{wall}}^{GF2}=2\text{hr}$ while for $\text{H}_{100}$
we have $T_{\text{wall}}^{MP2}=3.5\text{min}$ and $T_{\text{wall}}^{GF2}=40{\rm min}$. 

Note that these timings are for a single calculation. The error estimation
uses, as mentioned, ten completely independent runs, and therefore
took 10 times longer. 

For comparison, we note that the CPU time for the deterministic calculation
in the $\text{H}_{100}$ system takes $45$ min. on a single core,
which is 4 times faster than the stochastic calculation. Since the
deterministic algorithm scales steeply as $O\left(N^{5}\right)$,
the crossover occurs already at $\text{H}_{150}$ and at $\text{H}_{1000}$
the deterministic calculation would take $\frac{10^{4}-10^{5}}{n_{CORES}}$
wall time hours per SCF iteration, compared to $\frac{24}{n_{CORES}}$
hours for the stochastic calculations. 

\subsection{Born Oppenheimer potential curves}

\begin{figure}
\includegraphics[width=1\columnwidth]{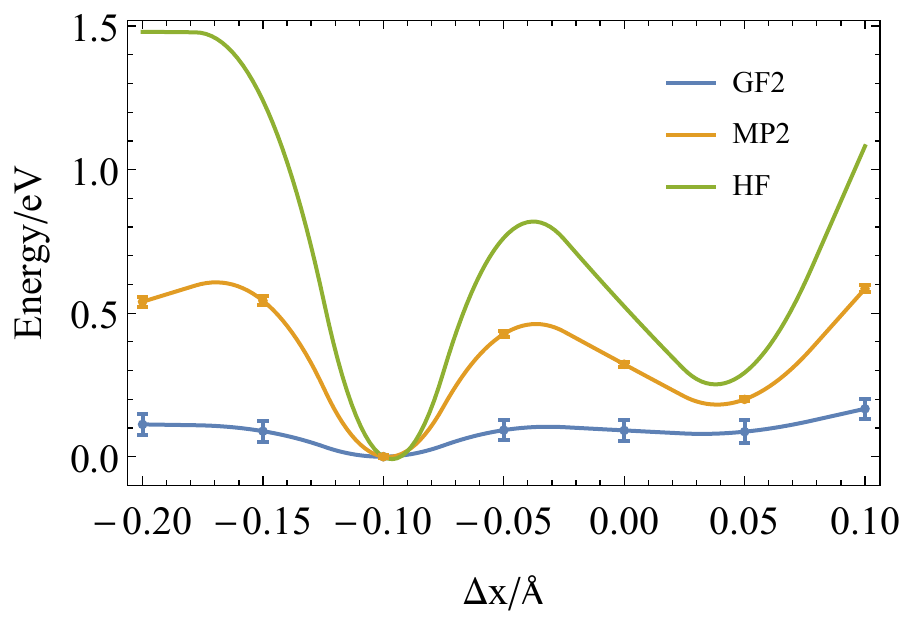}

\caption{\label{fig:PES}Hartree-Fock, MP2 and GF2 potential energy curves
for the energy change in the displacement $\Delta x$ of atom no.
25 in the 100 H chain.}
\end{figure}

Potential energy curves can be calculated by correlated sampling,
where at each new nuclear configuration one employs the same set of
stochastic orbitals $\eta_{0},$$\,\xi_{0}$ and $\zeta_{0}$ for
the self-energy estimation. For demonstration, the HF, MP2 and GF2
Born Oppenheimer potentials of the $\text{H}_{100}$ system are shown
in Fig.~\ref{fig:PES} as a function of the displacement of atom
no. 25 (counting from the left). In all three methods the most stable
position of the atom is at $\sim-0.1a_{0}$, slightly displaced towards
the nearest chain end. HF theory produces an energy potential with
large variations of up to $1.5$eV and large vibrational frequencies
of order of 3.4eV. The MP2 curve is much smoother and the vibrational
frequencies reduces to $\sim2.4$eV while the GF2 energy curve is
considerably flatter, predicting a vibrational frequency of $\sim1.0$
eV. 

\section{Summary and Conclusions}

The problem we addressed here is the reduction of the the steep ${\rm O({\it N}^{5})}$
scaling associated with the implementation of self-consistent GF2
calculations. We developed an effective way to reduce complexity to
${\rm O({\it N}^{3})}$ by using stochastic techniques for calculating
the self-energy. A detailed derivation was given along with a specific
algorithm. The sampling error in the overall algorithm was studied
for linear $\text{H}_{N}$ systems, and the simulation showed that
the stochastic errors in the correlation energies can be controlled
to less than 0.1\% for very large systems. While the studied systems
were linear, the algorithm makes no use of the linearity and applies
equally well to any geometry.

As a byproduct, since the first step in GF2 is equivalent to MP2,
we obtain a stochastic MP2 method (sMP2) performed on top of an existing
HF calculation. This approach too has a formal complexity of ${\rm O\!}\left(N^{5}\right)$
which is reduced here to \emph{linear} ${\rm O({\it N})}$, except
for a single overall Fock-matrix diagonalization which is often available
from the underlying HF or DFT ground-state calculation. The errors
in this well-scaling stochastic MP2 method are comparable to those
of local MP2 approaches used in quantum chemistry. 

For GF2, the method has two main stages. The first stage, as in the
MP2 case, is a linear scaling calculation of the self-energy. This
self-energy is then used in the second stage to construct the Green's
function, at an ${\rm O\!}\left(N^{3}\right)$ cost. A complication
arises in GF2 due to this second stage (but not in MP2!), where the
self-energy enters non-linearly into the expression for the Green's
function. This non-linearity gives rise to a noticeable bias which
is proportional to the system size $N$. To overcome this bias the
number of stochastic orbitals $I$ used in the first step must be
increased in proportion to the system size $N$, and hence the self-energy
calculation in GF2 attains an ${\rm O\!}\left(N^{2}\right)$ scaling.
The overall scaling of the GF2 calculation is unaffected by this bias
problem and remains ${\rm O\!}\left(N^{3}\right)$.

The present calculations give a fully self-consistent Green's function
method for a large system with a thousand electrons described by a
full quantum chemistry Hamiltonian. Moreover, we demonstrated that
the splitting of matrices by a random average over stochastically
chosen vectors leads to small variance and that relatively few Monte
Carlo samples already yield quite accurate correlation energies. The
reason for this excellent sampling dependence is two-fold: the stochastic
sampling inherently acts only in the space of atomic orbitals while
the actual spatial integrals (Eq. (\ref{eq:vHartree})) are evaluated
using a deterministic, numerically exact calculation; in addition,
since the Green's function matrices are smooth in imaginary time,
different random vectors can be used at each imaginary-time point
thereby enhancing the stochastic sampling efficiency.

We have shown that both sMP2 and sGF2 are suitable for calculating
potential energy curves or surfaces. Interestingly, for the $\text{H}_{N}$
systems the potential curve is much smoother and flatter than in HF
or MP2. 

As for future applications, we note that sGF2 and sMP2 methods are
automatically suitable for periodic systems, as all the deterministic
steps and the time-frequency transforms are very efficient when done
in the reciprocal $(k)$ space. The only additional detail is that
in periodic systems one needs to choose the random vectors to be in
$k$-space and then convert them to real-space, as detailed in an
upcoming article.

Finally, we also note that, beyond the results presented here, it
should also be possible to achieve further reduction of the stochastic
error with an embedded fragment approach, analogous to self-energy
embedding approaches, where a deterministic self-energy is calculated
for embedded saturated fragments as introduced for stochastic DFT
applications. \cite{Neuhauser2014a,arnon2017equilibrium}

Discussions with Eran Rabani are gratefully acknowledged. D.N. was
supported by NSF grant DMR-16111382, R.B. acknowledges support by
BSF grant 2015687 and D.Z. was supported by NSF grant CHE-1453894.
\providecommand{\latin}[1]{#1}
\makeatletter
\providecommand{\doi}
  {\begingroup\let\do\@makeother\dospecials
  \catcode`\{=1 \catcode`\}=2\doi@aux}
\providecommand{\doi@aux}[1]{\endgroup\texttt{#1}}
\makeatother
\providecommand*\mcitethebibliography{\thebibliography}
\csname @ifundefined\endcsname{endmcitethebibliography}
  {\let\endmcitethebibliography\endthebibliography}{}

\end{document}